\pdfoutput=1
\documentclass[final,5p,times,twocolumn]{elsarticle}

\usepackage{amssymb}
\usepackage{lineno}

\journal{NIM A}

\begin{document}

\begin{frontmatter}

\title{Performance of SiPMs in the nonlinear region}

\author{Jaime Rosado}
\ead{jrosadov@ucm.es}
\address{Departamento de F\'{i}sica At\'{o}mica, Molecular y Nuclear and UPARCOS\\
Universidad Complutense de Madrid, 28040 Madrid, Spain}

\begin{abstract}
Silicon photomultipliers present saturation effects as they have a limited number of pixels and work in Geiger mode.
Their response to light pulses in the nonlinear region is very complex for two reasons: pixel recharging after an avalanche affects the
trigger probability and charge multiplication of subsequent avalanches, and non-trivial effects due to crosstalk and
afterpulsing.

A parametrization of the nonlinear response of silicon photomultipliers was developed where the above effects were
readily accounted for. The model was tested on a setup of $\gamma$-ray spectrometry using different combinations of
scintillation crystals and detectors. The model parameters were interpreted in terms of fundamental characteristics of
the setup (e.g., lifetime of the scintillation crystal and pixel recovery time). The proper conversion from signal
resolution to energy resolution was provided.
\end{abstract}


\end{frontmatter}


\section{Introduction}
\label{sec:intro}

Silicon photomultipliers (SiPMs) are increasingly becoming the best-choice high-sensitivity photodetectors  for many
applications. However, one of their main drawbacks is that they go nonlinear at relatively low light intensity as a
consequence of their design concept. The detector output signal is the sum of the signals from all its pixels, where
each pixel is a Geiger-mode avalanche photodiode that produces signals of amplitude independent of the number of
photons hitting it at a time.

In the case of infinitely short light pulses and in the absence of other effects, the average output signal of a SiPM
is proportional to the mean number of fired pixels $\overline{N}_{\rm
fired}$, which is given by the well-known formula:
\begin{equation}
\label{eq:simple_model}
\overline{N}_{\rm fired}=N_{\rm pix}\cdot\left[1-\exp\left(\frac{-{\rm PDE}\cdot N_{\rm ph}}{N_{\rm pix}}\right)\right]\,,
\end{equation}
where $N_{\rm pix}$ is the total number of pixels of the SiPM, $N_{\rm ph}$ is the number of incident photons and PDE
is the photon detection efficiency. This expression deviates from linearity by $5\%$ when ${\rm PDE}\cdot N_{\rm ph}$ is only $10\%$ of $N_{\rm pix}$
(e.g., $N_{\rm ph}= 144$ for a typical $3\times3$~mm$^2$ SiPM with pixel pitch of 50~$\mu$m and ${\rm PDE}=40\%$).

At the moment an avalanche occurs in a pixel, the overvoltage (i.e., the excess voltage over breakdown voltage) on the
pixel drops to zero and then recovers exponentially. Subsequent avalanches can be produced in the pixel while
recharging, but with a lower probability and gain depending on the instantaneous overvoltage on the pixel. Therefore,
saturation effects for light pulses of duration comparable to the pixel recovery time $\tau_{\rm rec}$ should not be so
large as predicted by expression (\ref{eq:simple_model}). In addition, the SiPM response can no longer be described by
a static PDE and gain figures.

Another important issue of SiPMs is that they have correlated noise, that is, an avalanche triggered at a certain pixel
may induce secondary avalanches in neighboring pixels (crosstalk) or in the same pixel with some delay (afterpulsing).
These effects basically increase the response of the SiPM per incident photon and thus speed up saturation, but
modeling them in detail is complicated \cite{Rosado}.

A model of the SiPM response should take into account these features. As the problem is very complex, a working
solution is to use a simple ansatz that fits well to data (see, e.g., \cite{Niu}). On the other hand, this approach
does not provide an adequate understanding of the detector performance. Some authors have developed statistical models
that include the above-described stochastic processes in a comprehensive way \cite{vanDam,Vinogradov}. However, the
intricacies and interdependencies of these processes make necessary to adopt not obvious approximations that limit the
applicability and accuracy of these models.

In this work, I followed a different approach consisting of formulating an appropriate parametrization of the above
processes without the aim of modeling them in detail. The result is a simple expression that fit to experimental data
with only two free parameters, which have precise physical meanings and are clearly related to the characteristics of
the detector.

The model has been applied to and validated with measurements performed in a setup of $\gamma$-ray spectrometry using
different scintillation crystals and SiPMs. This allowed me to make the energy calibration of the setup and apply the
proper conversion from signal resolution to energy resolution.

\section{The model}
\label{sec:model}

As a proof of concept, the model was developed for the case that a SiPM detects a light pulse from a scintillation
crystal when a $\gamma$ ray with energy $E_\gamma$ undergoes a photoelectric interaction in it. In this case, the
number of incident photons $N_{\rm ph}$ on the SiPM sensitive area can be assumed to be a Poisson random variable with
mean $\overline{N}_{\rm ph}=G\cdot L\cdot E_\gamma$, where $G$ is the light collection efficiency and $L$ is the
luminosity of the scintillator.

\subsection{Number of avalanches}
\label{ssec:avalanches}

First, let us define $N_{\rm det}\equiv{\rm PDE}\cdot N_{\rm ph}$, hereafter referred to as number of ``detected
photons''. $N_{\rm det}$ is also a Poisson random variable and represents the number of avalanches that would be
triggered in an ideal SiPM with neither saturation effects or correlated noise. The next step is to relate $N_{\rm
det}$ to the actual number of triggered avalanches $N_{\rm av}$ including both losses of detected photons in charging
pixels and secondary avalanches due to crosstalk and afterpulsing. Dark noise was neglected for simplicity.

The probability distribution of $N_{\rm av}$ for the detection of a $\gamma$ ray of energy $E_\gamma$ can be expressed
as
\begin{equation}
\label{eq:prob_Nav}
P\left(N_{\rm av}=n\right)=\sum_{m=0}^\infty \frac{\lambda^m\cdot {\rm e}^{-\lambda}}{m!}\cdot P_m\!\left(n\right)\,,
\end{equation}
where $\lambda={\rm PDE}\cdot G\cdot L\cdot E_\gamma$ and $P_m\!\left(n\right)$ stands for the conditional probability
of triggering exactly $n$ avalanches given $m$ detected photons. In principle, this probability should be calculated
considering all the possible spatial patterns of fired pixels and the time sequences of avalanches in them, including
those due to crosstalk and afterpulsing. However this is only feasible by means of a Monte Carlo simulation.

Nevertheless, an approximate solution of $P_m\!\left(n\right)$ is given by the iterative formula:
\begin{equation}
\label{eq:iter_Nav}
P_{m+1}\!\left(n\right)\approx P_{m}\!\left(n\right)\cdot \frac{n}{N_{\rm sat}}
+\sum_{i=1}^n\!P_{m}\!\left(n-i\right)\cdot\left(1-\frac{n-i}{N_{\rm sat}}\right)\cdot P_{1}\!\left(i\right)\,,
\end{equation}
which was derived under the following assumptions:
\begin{enumerate}[$\bullet$]
\item The probability that the detected photon added in the $(m+1)$-th iteration is lost, i.e., it fires no pixel, is
    assumed to be proportional to the number of avalanches in previous iterations. This approximation can be made
    if the time distribution of triggered avalanches is not significantly distorted by saturation effects or
    correlated noise.
\item If the $(m+1)$-th photon is to fire a pixel, the probability distribution of the total number avalanches induced
    by this single photon (i.e., the primary one plus the secondary ones) is assumed to be unaffected by avalanches
    of previous iterations, if any. Therefore, $P_{1}\!\left(i\right)$ is used for every photon added iteratively.
\end{enumerate}

The coefficient $N_{\rm sat}^{-1}$ can be interpreted as the probability that, for any pair of detected photons, they
both hit the same pixel and the avalanche triggered by the first photon prevents the second one from triggering another
avalanche. An estimate of this probability can be obtained under the assumptions that the breakdown probability is
proportional to the instantaneous overvoltage on the pixel and that the scintillation pulse is described by a pure
exponential distribution with time constant $\tau$ (i.e., the scintillation decay time). This leads to:
\begin{equation}
\label{eq:Nsat}
N_{\rm sat}\approx N_{\rm pix}\cdot\left(1+\frac{\tau}{\tau_{\rm rec}}\right)\,.
\end{equation}
Note that $N_{\rm sat}$ is always greater than $N_{\rm pix}$. The ratio $\tau/\tau_{\rm rec}$ is a measure of how close
in time two photons are on average. In the limit $\tau\gg\tau_{\rm rec}$, it is obtained $N_{\rm sat}^{-1}\to0$, which
means that losses of detected photons are negligible.

From expressions (\ref{eq:prob_Nav}) and (\ref{eq:iter_Nav}), the average of $N_{\rm av}$ is
\begin{equation}
\label{eq:mean_Nav}
\overline{N}_{\rm av}\approx N_{\rm sat}\cdot\left[1-\exp\left(\frac{-{\rm PDE}\cdot E_\gamma}{E_{\rm sat}}\right)\right]\,,
\end{equation}
where $E_{\rm sat}$ has units of energy and is defined as
\begin{equation}
\label{eq:Esat}
E_{\rm sat}^{-1}=E_{\rm sat,0}^{-1}\cdot \sum_{i=1}^\infty i\cdot P_{1}\left(i\right)\,,
\end{equation}
with $E_{\rm sat, 0}^{-1}\equiv G\cdot L\cdot N_{\rm sat}^{-1}$ being a constant. That is, $E_{\rm sat}^{-1}$ is
proportional to the average number of avalanches induced by a single detected photon, which depends on $\Delta V$. In
the limit $\Delta V\to 0$, the probability of correlated noise is zero and $E_{\rm sat}^{-1}\to E_{\rm sat,0}^{-1}$.

In principle, in equation (\ref{eq:mean_Nav}), the PDE should be averaged on the light emission spectrum of the scintillator.
Nevertheless, without loss of generality, $E_{\rm sat}$ could be defined to include the PDE or, at least, the part of it
that is not accounted for when using handy reference PDE data (e.g., at a given wavelength). In addition, for other
applications, the above definitions can be adapted to use the amplitude of the light pulse or some other convenient
magnitude instead of $E_\gamma$.

The expression (\ref{eq:mean_Nav}) resembles expression (\ref{eq:simple_model}), but with two important differences: i)
the saturation level $N_{\rm sat}$ is higher than $N_{\rm pix}$, since each pixel can be fired several times per pulse,
ii) the saturation rate, determined by $E_{\rm sat}$, includes the contribution of correlated noise. Since several
approximations were made to obtain equation (\ref{eq:mean_Nav}), determining $N_{\rm sat}$ and $E_{\rm sat}$ in terms
of elemental parameters is useless. Instead, it is more convenient to let $N_{\rm sat}$ and $E_{\rm sat}$ as free
parameters to be estimated from experimental data, as explained below.

\subsection{Output charge}
\label{ssec:charge}

The average output charge $\overline{Q}$ for a $\gamma$ ray of energy $E_\gamma$ is $\overline{N}_{\rm av}$ times the
mean avalanche charge including gain losses in recharging pixels. Similarly to the losses of detected photons, these
gain losses can be assumed to be proportional to the number of avalanches. Assuming also $\sigma_{N_{\rm
av}}\ll\overline{N}_{\rm av}$ for a scintillation pulse, $\overline{Q}$ is approximately given by
\begin{equation}
\label{eq:mean_Q}
\overline{Q}\approx \overline{N}_{\rm av}\cdot Q_1\cdot\left(1-\frac{\beta\cdot \overline{N}_{\rm av}}{N_{\rm sat}}\right)\,.
\end{equation}
where $\beta$ is other model parameter and $Q_1$ stands for the mean avalanche charge in a steady pixel, which can be
measured from the output charge spectrum for single photons and is proportional to $\Delta V$.

Note that $N_{\rm sat}$ and $\beta$ are strongly correlated to each other because both parameters should be basically
determined by the ratio $\tau/\tau_{\rm rec}$. Making the same assumptions as for equation (\ref{eq:Nsat}), the
following relationship between $N_{\rm sat}$ and $\beta$ was deduced:
\begin{equation}
\label{eq:beta}
\beta\approx\frac{1-N_{\rm pix}/N_{\rm sat}}{2-N_{\rm pix}/N_{\rm sat}}\,.
\end{equation}

Equation (\ref{eq:mean_Q}), in combination with equations (\ref{eq:mean_Nav}) and (\ref{eq:beta}), allows one to describe
$\overline{Q}$ as a function of $E_\gamma$ at each $\Delta V$ value. Fitting this function
$f\left(E_\gamma\right)$ to a set of $\overline{Q}$ data measured at several $E_\gamma$ and $\Delta V$ values provides a
complete energy calibration of the detector response. The only two free parameters in this fit are $N_{\rm sat}$ and
$E_{\rm sat}$, where $E_{\rm sat}$ should be let to vary with $\Delta V$.

\subsection{Energy resolution}
\label{ssec:resolution}

Once the detector is calibrated, any measured charge output $Q$ can be related to a certain energy $E_{\rm
cal}=f^{-1}(Q)$, such that the central energy associated to a photopeak is $\overline{E}_{\rm
cal}=f^{-1}(\overline{Q})=E_\gamma$. Therefore, if $R_Q$ is the charge resolution measured on the photopeak, the energy
resolution of the detector at that energy is simply given by:
\begin{equation}
\label{eq:E_resolution1}
R_{E_{\rm cal}}=\frac{\overline{Q}/E_\gamma}{f'\left(E_\gamma\right)}\cdot R_Q\,.
\end{equation}
Solving the derivative of $f\left(E_\gamma\right)$, this leads to
\begin{equation}
\label{eq:E_resolution2}
R_{E_{\rm cal}}=\frac{E_{\rm sat}}{{\rm PDE}\cdot E_\gamma}\cdot\frac{\overline{N}_{\rm av}}{N_{\rm sat}-\overline{N}_{\rm av}}
\cdot\frac{\overline{Q}}{2\cdot\overline{Q}-Q_1\cdot\overline{N}_{\rm av}}\cdot R_Q\,.
\end{equation}
It can be easily seen that this relationship reduces to $R_{E_{\rm cal}}= R_Q$ in the linear limit ${\rm PDE}\cdot
E_\gamma\ll E_{\rm sat}$.

\section{Experimental validation}
\label{sec:experiment}

\subsection{Method}
\label{ssec:method}

Measurements were performed for several combinations of four different SiPMs and three scintillon crystals. The chosen
SiPMs (see table \ref{tab:SiPMs}) were of the S13360 series from Hamamatsu \cite{Hamamatsu}, which have very low
crosstalk and afterpulsing. For application of the model, I used data of PDE as a function of $\Delta V$ provided by
the manufacturer for 405~nm photons.

\begin{table}[!b]
\caption{Characteristics of the chosen SiPMs. Data of $\tau_{\rm rec}$ were taken from \cite{Rosado}.}%
\label{tab:SiPMs}%
\smallskip%
\centering%
\begin{tabular}{|ccccc|}
\hline
            & Sensitive area & Pixel pitch & $N_{\rm pix}$ & $\tau_{\rm rec}$ \\
S13360  & (mm$^2$)       & ($\mu$m)    &               & (ns)             \\
\hline
-1325CS & $1.3\times1.3$ & 25          &  667          & 17               \\
-3025CS & $3.0\times3.0$ & 25          & 3600          & 12               \\
-1350CS & $1.3\times1.3$ & 50          &  667          & 29               \\
-3050CS & $3.0\times3.0$ & 50          & 3600          & 29               \\
\hline
\end{tabular}
\end{table}

The materials of the three scintillation crystals were LYSO, LFS and CsI(Tl), which have high luminosities ranging from
30 to 60~photons/keV. The LYSO and LFS scintillators are characterized by a short decay time $\tau\sim40$~ns, while
$\tau\sim1\,\mu$s for CsI(Tl). All the crystals had a square base of $3\times3$~mm$^2$ so that they covered the full
sensitive area of any SiPM. Silicone grease was used for the optical coupling.

The SiPM bias voltage was regulated using a Hamamatsu C12332 driver circuit that includes temperature compensation
\cite{Hamamatsu}. Instead of employing conventional spectrometry instrumentation, the signal was registered by a
digital oscilloscope (Tektronix TDS5032B), since it offers a very wide dynamic range that allowed the recording of both
saturated signals and single-photon signals without the need of amplification. The output charge of the SiPM was
measured by integrating the signal over a time interval long enough to include the entire scintillation pulse.

The crystals were irradiated by different weak radioactive sources ($^{22}$Na, $^{60}$Co, $^{137}$Cs and $^{226}$Ra) to
characterize the detector response for an ample range of $\gamma$-ray energies from 300 to 2100~keV. The $\overline{Q}$
value of each identified photopeak in the measured spectra was obtained by Gaussian fitting. The well isolated photopeak due to $\gamma$ rays of 662~keV of the $^{137}$Cs spectrum was used to
characterize the detector energy resolution.

\subsection{Results}
\label{ssec:results}

\begin{figure}[t!]
\centering
\includegraphics[width=1\linewidth]{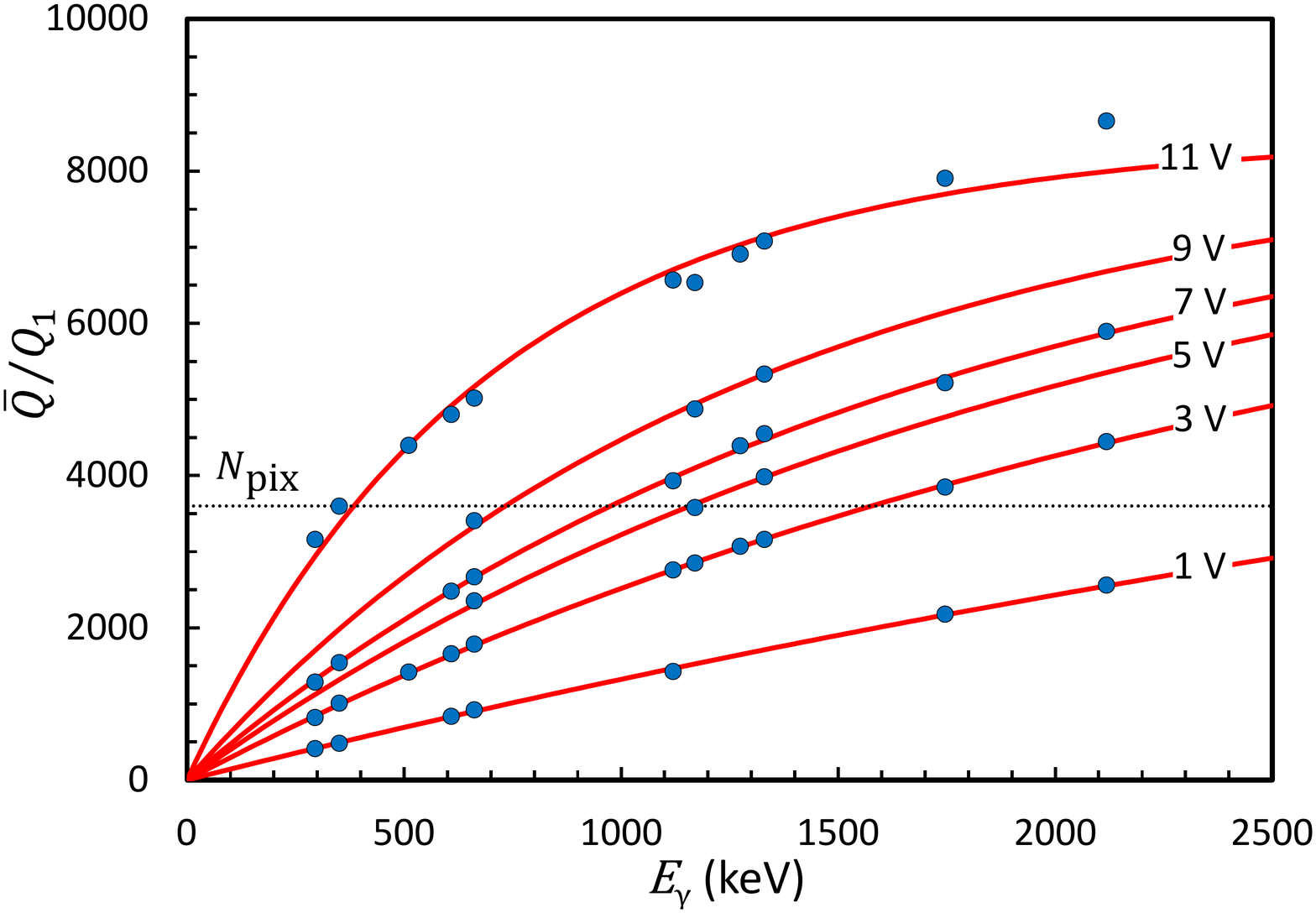}
\caption{Model fitting to data for the S13360-3050CS SiPM coupled to the LFS crystal.
The labels indicate the applied overvoltage.
The dotted horizontal line marks the saturation level predicted by equation (\ref{eq:simple_model}).}
\label{fig:Q_E}
\end{figure}

The model fits well to data at moderate saturation for all the tested SiPM-crystal combinations. As an example, figure
\ref{fig:Q_E} displays the detector response versus energy at several $\Delta V$ values for the S13360-3050CS SiPM
coupled to the LFS crystal, where nonlinearity is clearly seen. It is represented the normalized average charge
$\overline{Q}/Q_1$ for better visualization. The dotted horizontal line marks the value $N_{\rm pix}=3600$
corresponding to the saturation level of $\overline{Q}/Q_1$ in case that each pixel could only be fired once during a
scintillation pulse, as assumed by the crude approximation of expression (\ref{eq:simple_model}). Notice that the
detector response readily surpasses this level, meaning that the probability that a pixel is fired while charging is
high, since $\tau_{\rm rec}$ and $\tau$ are comparable for this SiPM-crystal combination. The fit yields $N_{\rm
sat}=15100$, therefore the model prediction is that $\overline{Q}/Q_1$ saturates at $N_{\rm sat}\cdot(1-\beta)=8600$.
However, as commented before, the model is expected to fail near saturation. Indeed, experimental data at high
overvoltage and large energy overpass this limit.

Results for other SiPM-crystal combinations are not shown, but some remarks are given next. Both the S13360-1350CS and
S13360-3050CS SiPMs, with a pixel pitch of 50~$\mu$m, exhibited noticeable nonlinearity when coupled to the LYSO and
LFS crystals, whereas they behaved almost linearly in the whole range of test energies when coupled to the CsI(Tl)
crystal. This is due to the fact that $\tau\gg\tau_{\rm rec}$ for the CsI(Tl) scintillator. Nonlinearity is less
important in the S13360-1325CS and S13360-3025CS SiPMs as expected from their smaller pixel pitch and shorter
$\tau_{\rm rec}$. In general, the ratio $N_{\rm sat}/N_{\rm pix}$ was found to be approximately proportional to
$1+\tau/\tau_{\rm rec}$, as predicted by equation (\ref{eq:Nsat}). For example, $N_{\rm sat}/N_{\rm pix}=54$ for both
the S13360-1350CS and S13360-3050CS SiPMs when coupled to the CsI(Tl) crystal, and $N_{\rm sat}/N_{\rm pix}\sim5$ for
the LYSO and LFS crystals. For the S13360-1325CS and S13360-3025CS SiPMs coupled to the LYSO or LFS crystals, $N_{\rm
sat}/N_{\rm pix}\sim9$.

\begin{figure}[t!]
\centering
\includegraphics[width=1\linewidth]{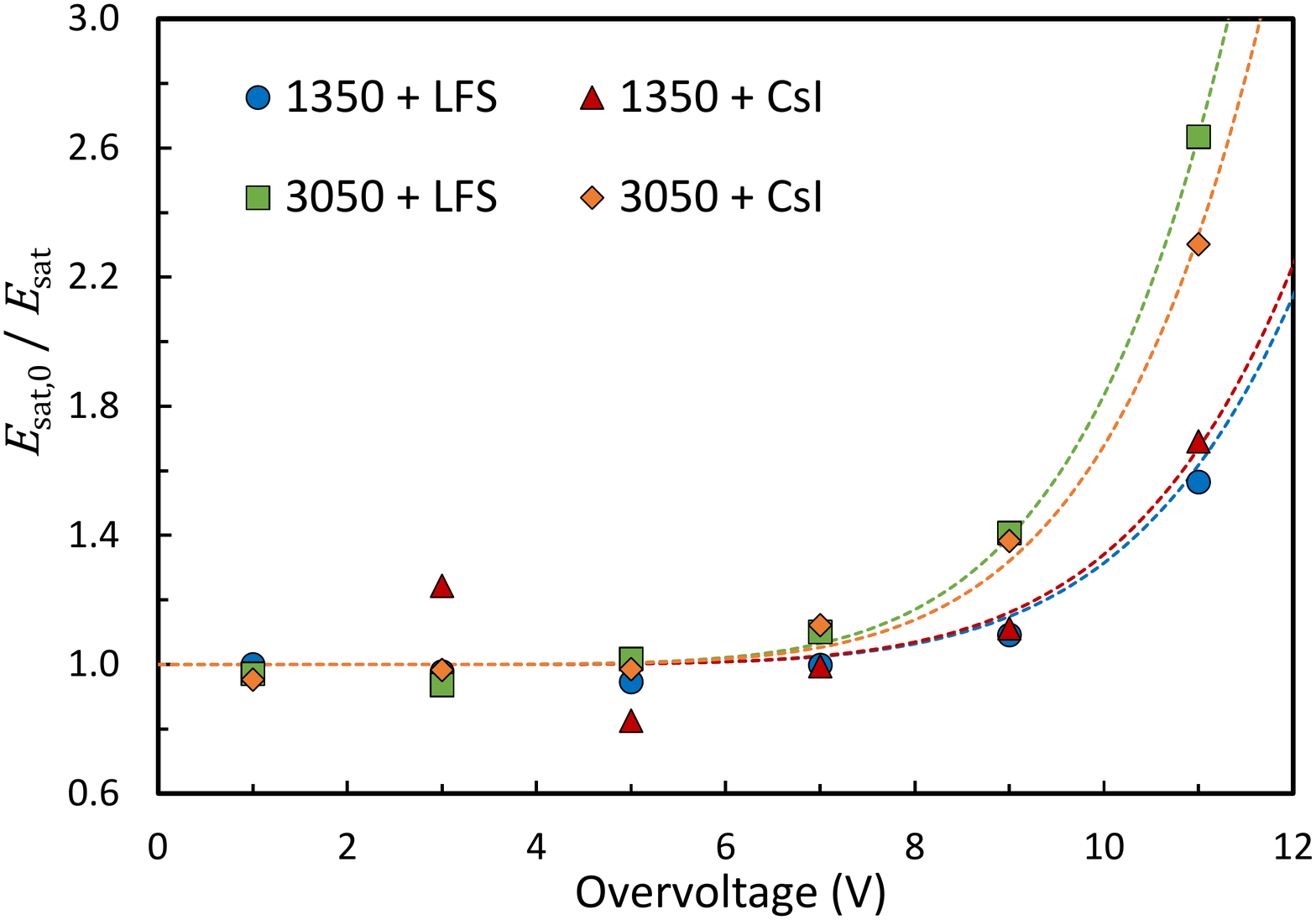}
\caption{Normalized $E_{\rm sat}^{-1}$ values as a function of overvoltage for several combinations of SiPMs and scintillation crystals.
A power law fits well to data (dotted curves), as expected for the behavior of correlated noise.
It is clearly seen that the S13360-3050CS SiPM presents more correlated noise that the S13360-1350CS one, in agreement with results from \cite{Rosado}.}
\label{fig:Esat}
\end{figure}

The fitted $E_{\rm sat}$ values for some SiPM-crystal combinations are shown in figure \ref{fig:Esat} as a function of
applied overvoltage. The ratio $E_{\rm sat,0}/E_{\rm sat}$ is displayed instead for comparison purposes, as this ratio
only depends on the correlated noise. Aside from statistical fluctuations, a power law (dotted lines in the figure)
describes well data. It can be clearly seen that the S13360-3050CS SiPM presents more correlated noise that the
S13360-1350CS SiPM, in agreement with results from \cite{Rosado}.

\begin{figure}[t!]
\centering
\includegraphics[width=1\linewidth]{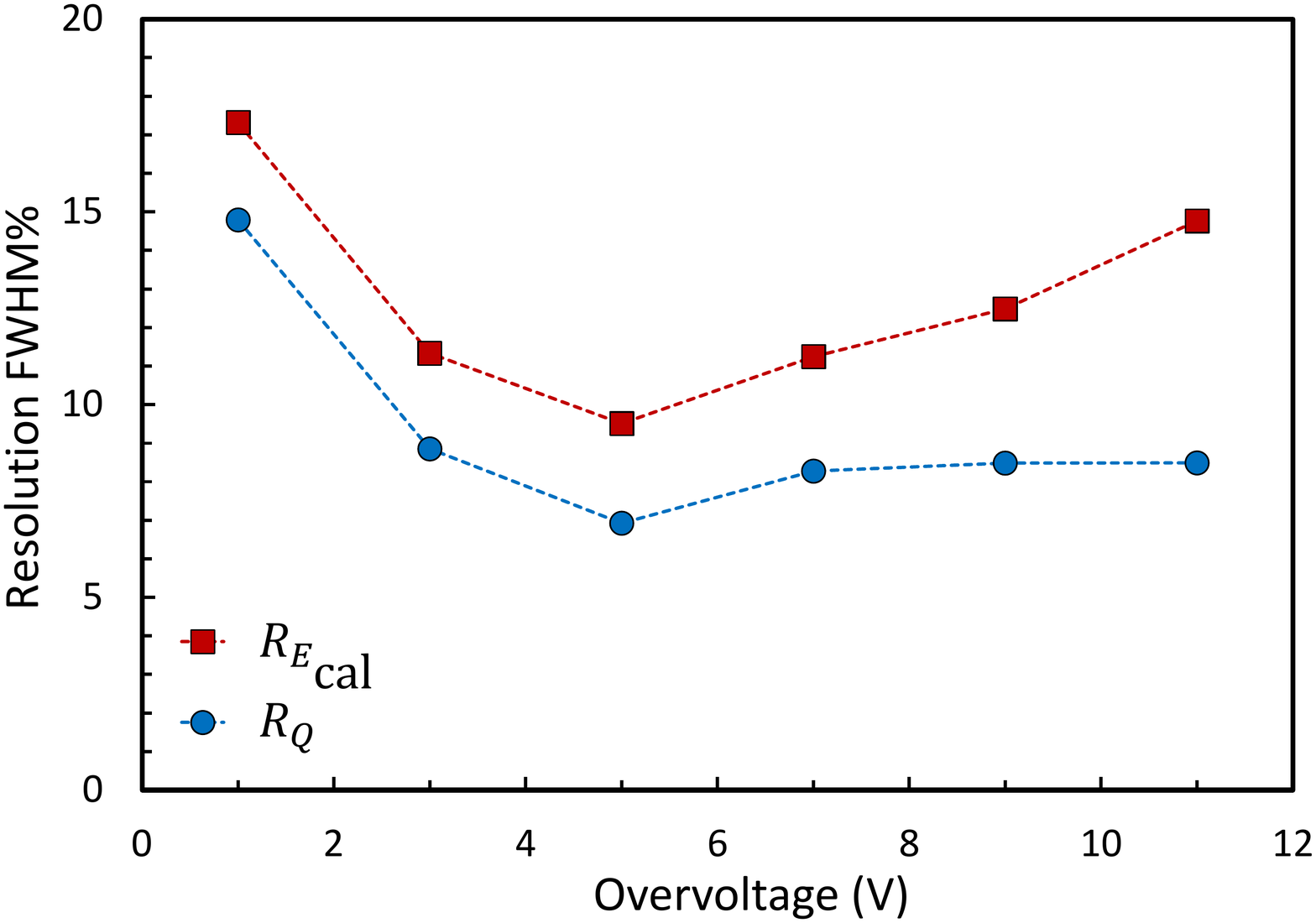}
\caption{$R_Q$ and $R_{E_{\rm cal}}$ at 662~keV for the S13360-1350CS SiPM coupled to the LYSO crystal.
The conversion from $R_Q$ into $R_{E_{\rm cal}}$ was made using equation (\ref{eq:E_resolution2}).
$R_Q$ grows smoothly with overvoltage, but $R_{E_{\rm cal}}$ increases quickly.}
\label{fig:Resolution}
\end{figure}

An example of the $R_Q$ and $R_{E_{\rm cal}}$ values obtained at 662~keV for the S13360-1350CS SiPM coupled to the LYSO
crystal is in figure \ref{fig:Resolution}. The measured $R_Q$ reaches a minimum of 6.9\% (FWHM) at $\Delta V=5$~V and
then grows smoothly. The $R_{E_{\rm cal}}$ curve was obtained from $R_Q$ using the proper conversion formula
(\ref{eq:E_resolution2}). $R_{E_{\rm cal}}$ is higher (e.g., 9.2\% at 5~V) and grows much faster with
overvoltage as a consequence of the increasing saturation effects. Similar $R_{E_{\rm cal}}$ values of the order of
10\% were obtained for other SiPM-crystal combinations, but with a less pronounced increase of $R_{E_{\rm cal}}$ with
$\Delta V$ in the cases that saturation effects are low.

\section{Conclusions}
\label{sec:conclusions}

A model was developed to describe the nonlinear response of SiPMs to light pulses using only two fitting parameters. It
accounts for the fact that pixels can be fired multiple times during a light pulse, but both their average breakdown
probability and gain are reduced during pixel recharging. Effects from crosstalk and afterpulsing depending on bias voltage are
also included.

The model was applied to and validated with measurements of $\gamma$-ray spectrometry. Experimental results were
interpreted with the aid of the model in terms of the characteristics of the SiPM and the scintillation crystal (e.g.,
pixel recovery time and scintillation decay time). The model also provided the correct way to convert the output charge
resolution into energy resolution.

\section*{Acknowledgements}
This work was supported by the Spanish MINECO under contract FPA2015-69210-C6-3-R.

\end{document}